\documentclass[jpsj_ref]{jpsj2}
\usepackage{graphicx}
\usepackage{bm}
\title{
Threshold Current of Domain Wall Motion under Extrinsic Pinning, 
$\beta$-Term and Non-Adiabaticity
}
\author{
Gen Tatara$^{1,2}$,
Toshihiko Takayama$^{1}$,
Hiroshi Kohno$^{3}$,
Junya Shibata$^{4}$,
Yoshinobu Nakatani$^{5}$
and
Hidetoshi Fukuyama$^{6}$
}

\inst{
$^1$Graduate School of Science, Tokyo Metropolitan University\\
1-1 Minamiosawa, Hachioji,  Tokyo 192-0397\\
$^2$PRESTO, JST, 4-1-8 Honcho Kawaguchi, Saitama 332-0012\\
$^3$
Graduate School of Engineering Science, Osaka University,
Toyonaka, Osaka 560-8531\\
$^4$
Frontier Research System (FRS), The Institute of Physical and
Chemical Research (RIKEN), 2-1 Hirosawa, Wako, Saitama 351-0198\\
$^5$
Department of Computer Science
The University of Electro-Communications,
1-5-1 Chofugaoka, Chofu-shi,
Tokyo, 182-8585\\
$^6$
International Frontier Center for Advanced Materials\\
Institute for Materials Research, Tohoku University\\
2-1-1 Katahira, Aoba-ku, Sendai 980-8577
}

\abst{
Threshold current of domain wall motion under spin-polarized electric current in ferromagnets is theoretically studied based on the equation of motion of a wall in terms of collective coordinates.
Effects of non-adiabaticity and a so-called $\beta$-term in Landau-Lifshitz equation, 
which are described by the same term in the equation of motion of a wall, 
are taken into account as well as extrinsic pinning.
It is demonstrated that there are four different regimes
 characterized by different dependence of threshold on extrinsic pinning, hard-axis magnetic anisotropy, non-adiabaticity and $\beta$.
}

\kword{
spin torque, domain wall, threshold current, MRAM
}

\begin{document}

\maketitle
\def\average#1{\langle {#1} \rangle}
\newcommand{\betac}{{\beta^{\rm c}}}
\newcommand{\betap}{{\beta'}}
\newcommand{\Bc}{B_{\rm c}}
\newcommand{\jc}{j_{\rm c}}
\newcommand{\ja}{{j^{\rm a}}}
\newcommand{\js}{{j_{\rm s}}}
\newcommand{\jsv}{{\jv_{\rm s}}}
\newcommand{\jci}{j^{\rm i}_{\rm c}}
\newcommand{\jce}{j^{\rm c(ex)}}
\newcommand{\jatil}{{\tilde{j}_{\rm a}}}
\newcommand{\jctil}{{\tilde{j}_{\rm c}}}
\newcommand{\jtil}{{\tilde{j}}}
\newcommand{\Kp}{{K_\perp}}
\newcommand{\Mw}{{M_{\rm w}}}
\newcommand{\muB}{\mu_B}
\newcommand{\om}{{\omega}}
\newcommand{\Omegap}{{{\Omega'}}}
\newcommand{\Omegatil}{{\tilde{\Omega}}}
\newcommand{\Omegaptil}{{\tilde{\Omegap}}}
\newcommand{\Ptil}{{\tilde{P}}}
\newcommand{\Rw}{{R_{\rm w}}}
\newcommand{\sgn}{{\rm sgn}}
\newcommand{\tauv}{\bm{\tau}}
\newcommand{\tautil}{{\tilde{\tau}}}
\newcommand{\ttil}{{\tilde{t}}}
\newcommand{\vc}{{v_{\rm c}}}
\newcommand{\ve}{{v_{\rm e}}}
\newcommand{\Vz}{{V_0}}
\newcommand{\Vztil}{{\tilde{V_0}}}
\newcommand{\Xtil}{{\tilde{X}}}
\newcommand{\av}{{\bm a}}
\newcommand{\bv}{{\bm b}}
\newcommand{\betatil}{{\tilde{\beta}}}
\newcommand{\cv}{{\bm c}}
\newcommand{\chiuni}{{\chi_{0}}}
\newcommand{\DOS}{{\nu}}
\newcommand{\ells}{{\ell_{\rm s}}}
\newcommand{\Hp}{{H'}}
\newcommand{\jv}{{\bm j}}
\newcommand{\kv}{{\bm k}}
\newcommand{\Ne}{{N_{\rm e}}}
\newcommand{\nimp}{{n_{\rm imp}}}
\newcommand{\omegaell}{{\omega_{\ell}}}
\newcommand{\phiz}{{\phi}}
\newcommand{\rhow}{{\rho_{\rm w}}}
\newcommand{\Sv}{{\bm S}}
\newcommand{\taus}{{\tau_{\rm s}}}
\newcommand{\tr}{{\rm tr}}
\newcommand{\xv}{{\bm x}}
\newcommand{\Xv}{{\bm X}}
\newcommand{\vimp}{{v_{\rm imp}}}
\newcommand{\Tz}{\js}
\newcommand{\Tv}{{\bm T}}
\newcommand{\nv}{{\bm n}}
\newcommand{\mv}{{\bm m}}
\newcommand{\eF}{{\epsilon_F}}
\newcommand{\Ev}{{\bm E}}
\newcommand{\kF}{{k_F}}
\newcommand{\kp}{{k_\perp}}
\newcommand{\sigmav}{{\bm \sigma}}
\newcommand{\Mv}{{\bm M}}
\newcommand{\tilv}{{\tilde{v}}}
\newcommand{\tilK}{{\tilde{K}_\perp}}
\newcommand{\Area}{A}
\newcommand{\dx}{{d^3 x}}
\newcommand{\rv}{{\bm r}}
\newcommand{\qv}{{\bm q}}
\newcommand{\Vv}{{\bm V}}
\newcommand{\vv}{{\bm v}}
\newcommand{\Av}{{\bm A}}
\newcommand{\Bv}{{\bm B}}
\newcommand{\kB}{{k_B}}
%

\section{Introduction}

Intensive studies, both theoretical and experimental,
have been carried out to understand and control current-driven domain wall dynamics in nanoscale magnets
\cite{Grollier02,Vernier03,Tsoi03,Klaeui03,Yamaguchi04,Yamanouchi04,SMYT04,Yamanouchi06,
Berger78,Berger84,Berger92,TK04,TK05r,TK05r2,Thiaville04,Zhang04,Thiaville05,Barnes05,TVF05,STK05,Himeno05,Klaui05,Berger05,Ohe06,Tserkovnyak06,Maekawa06,Marrows05}.
So far all the experimental results\cite{Grollier02,Tsoi03,Klaeui03,Yamaguchi04,Yamanouchi04,Yamanouchi06} indicate that wall motion can really be induced by a current density larger than the threshold, $\jc$, and this threshold is of the order of $10^{12}$[A/m$^2$] in metals\cite{Grollier02,Tsoi03,Klaeui03,Yamaguchi04} and
 of $10^{9}$[A/m$^2$] in magnetic semiconductor\cite{Yamanouchi04}.
The key issue yet to be understood is the origin of this threshold and driving mechanism.

Theoretically, current-driven domain wall motion was predicted and studied by Berger since 1978\cite{Berger78,Berger84,Berger92}. He showed\cite{Berger92} that the motion in the limit of thick wall (adiabatic limit) is induced by the transfer of spin angular momentum from conduction electron to the wall via exchange interaction (spin transfer).
Another driving mechanism due to electron scattering (momentum transfer) was also pointed out on a phenomenological ground\cite{Berger84}.
Wall dynamics was discussed in terms of two variables, wall position $X$ and polarization (angle from the easy plane) $\phi$, 
namely,  within a rigid wall approximation introduced by Slonczewski\cite{Slonczewski72,Hubert98}.
Recently, this problem was reformulated  
from a microscopic point of view by two of the authors\cite{TK04}
within a rigid-wall approximation.
The work of Berger was extended there to systematically incorporate non-adiabaticity, and the momentum transfer effect was shown to be governed by the wall resistance, $\Rw$.  
A solution in the adiabatic limit in the absence of extrinsic pinning was presented there and existence of threshold current was pointed out\cite{TK04}. This threshold is due to an intrinsic pinning arising from a deformation (change of $\phi$), and the threshold was shown to be given by hard axis anisotropy (or demagnetization field), $\Kp$, as
\begin{equation}
 \jci =  \frac{e S^2}{a^3 \hbar P} \Kp \lambda,
\end{equation}
where $\lambda$ is wall thickness, 
$P\equiv \js/j$ is the polarization of current, 
$S$ and $a$ being the magnitude of local spin and lattice constant, respectively.

Recently a new torque term in Landau-Lifshitz-Gilbert (LLG) equation was proposed by Zhang and Li\cite{Zhang04} and Thiaville et al\cite{Thiaville05}.
This torque, proportional to $\beta \Sv\times (\jsv\cdot\nabla)\Sv$, where $\jsv$ is spin current and $\beta$ is a coefficient, is perpendicular to the spin transfer torque, $(\jsv\cdot\nabla)\Sv$, and is argued to arise from spin
relaxation of conduction electron\cite{Zhang04,Tserkovnyak06}, modification of damping by current\cite{Barnes05}, etc.\cite{Edwards05}
This term (sometimes called a $\beta$ term) was shown\cite{Zhang04,Thiaville05} to remove the threshold of the intrinsic origin, and that the wall starts to move for any small value of spin current 
with a finite velocity given by 
$\dot{X}\propto \frac{\beta}{\alpha}\js$, where $\alpha$ is Gilbert damping parameter, if extrinsic pinning is absent.
Smearing of intrinsic threshold occurs also due to non-adiabaticity.
This is seen from the fact that $\beta$ term plays exactly the same role as momentum-transfer effect on the domain wall dynamics described by $X$ and $\phi$ (i.e., as far as wall deformation besides $\phiz$-mode is neglected)\cite{Thiaville05}.
Existence of such term was shown theoretically in a different system of ferromagnetic junction\cite{Edwards05}.
At finite temperature, we have to take account of thermally assisted process, which also contributes to smoothen the velocity-current curve near the intrinsic threshold, $\jci$\cite{TVF05}.
In refs. \cite{Barnes05,Tserkovnyak06}, value of $\beta$ was discussed to be identical to $\alpha$ 
(in the case of a single band model in ref.\cite{Tserkovnyak06}).
In the present paper, we neverthelss treat $\beta$ and $\alpha$ as independent, since in reality, multiband effect\cite{Tserkovnyak06} and contribution to $\alpha$ from non-electron origin would result in a difference.
In the case of rigid domain wall we consider, $\beta$ is effectively replaced by $\betap$ below, which includes the effect of momentum transfer, and thus $\betap$ and $\alpha$ should be regarded as independent.

Once non-adiabaticity or the $\beta$ term are taken into account, pinning of extrinsic origin (such as defects) becomes essential in the dynamics, since
extrinsic pinning blocks the motion at low current.
This was indicated in a simulation by Thiaville et al\cite{Thiaville05}.
They simulated extrinsic pinning by surface roughness of the wire and showed that threshold current depends on roughness.
Ohe and Kramer\cite{Ohe06} recently studied a wall motion solving fully the torque arising from the conduction electron numerically.
Non-adiabaticity (momentum transfer force) was thus fully taken into account there. 
Extrinsic pinning being neglected, the threshold obtained was zero, consistent with the result of Thiaville et al, where the non-adiabaticity was effectively taken account of in terms of  $\beta$\cite{Thiaville05}.
These analysis so far are not enough to draw conclusion on the relation among threshold, $\beta$-term and extrinsic pinning. 
The aim of the present paper is to explore the wall dynamics with both $\beta$ term (and non-adiabaticity) and extrinsic pinning taken into account on a basis of the equation of motion of a domain wall.

In this paper, the sum of $\beta$ contribution and non-adiabaticity, i.e., the coefficient $\betap$ (Eq. (\ref{betap})), is assumed as positive.
Positive $\betap$ seems reasonable at present in that all the so far proposed origins of $\beta$, spin relaxation\cite{Zhang04} and 
modification of damping\cite{Barnes05,Tserkovnyak06}, leads to positive $\beta$ and 
the effect of non-adiabaticity (below Eq. (\ref{betap})) is also positive.
Some of the results in the weakly pinned regime are modified if $\betap <0$ as we mention below.

\section{Equation of motion of domain wall}

The equation of motion of local spin under current is written as 
\cite{Zhang04,Thiaville05}
\begin{equation}
\dot{\Sv}= \Bv_{\rm eff} \times \Sv
+\frac{\alpha}{S} \Sv\times\dot\Sv
-\frac{a^3}{2eS}(\jsv\cdot\nabla)\Sv
- \frac{a^3 \beta}{eS} [\Sv \times (\jsv\cdot\nabla)\Sv]
+\tauv_{\rm na}+\tauv_{\rm pin},
\label{modLLG}
\end{equation}
where $\Bv_{\rm eff}$ is the effective field arising from spin Hamiltonian, $\alpha$ represents Gilbert damping, which is introduced phenomenologically.
Spin torque (spin-transfer effect) is represented by a term  $(\jsv\cdot\nabla)\Sv$. 
This term is derived in the adiabatic limit\cite{Bazaliy98}, and thus the spin $\Sv$ in the above equation of motion must be slowly varying compared with Fermi wavelength, $k_{F}^{-1}$.
$\tauv_{\rm na}$ and $\tauv_{\rm pin}$ represent torque from the non-adiabaticity and pinning, respectively.
Explicit forms of these two terms are discussed when the equation of motion of domain wall is discussed. 
In fact, non-adiabaticity is not easily expressed in the equation of motion of a spin, since $\tauv_{\rm na}$ is non-local in space\cite{Kohno05}, while its effect on collective object such as domain wall is clearer as was discussed in ref. \cite{TK04}.

We consider a planar domain wall and take account of only a deformation mode described by the polarization angle $\phi$ out of the easy plane.
This is justified if the wire is narrow ($L_\perp \lesssim \lambda$, $L_\perp$ being the wire width) and
if the hard-axis anisotropy, $\Kp$, is weaker than the easy-axis one, $K$\cite{TT96}.
Such wall is described by the collective coordinates, $X$ and $\phi$, and the equation of motion derived from Eq. (\ref{modLLG}) reads\cite{TK04,Zhang04,Thiaville05}
\begin{eqnarray}
\frac{\dot{X}}{\lambda} -\alpha \dot{\phiz}
 &=& \frac{\vc}{\lambda} \sin 2\phiz + \frac{a^3}{2eS\lambda}Pj
 \nonumber\\
 \dot{\phiz} +\alpha\frac{\dot{X}}{\lambda}  
       &=& f_{\rm pin} +\betap \frac{a^3}{e\lambda} j ,
\label{DWeq}
\end{eqnarray}
where 
$\vc\equiv \Kp\lambda S/(2\hbar)$ corresponds to the drift velocity of the electron spin at $j=\jci$, and $P\equiv \js/j$ is the polarization of the current.
Left-hand side of the second equation represents total force acting on the wall. The force contribution from the current $j$ is a sum of $\beta$-contribution and non-adiabatic contribution (in other words, momentum-transfer effect)\cite{TK04};
\begin{equation}
\betap \equiv 
\beta P  +\beta_{\rm na},\label{betap}
\end{equation}
where
$\beta_{\rm na}\equiv \frac{\lambda}{2\hbar S}e^2 n\Rw A$ is a dimensionless measure of wall resistivity, 
$n$ is the electron density, $I\equiv j A$, $A$ being the crosssectional area of the wire.
Pinning potential is treated as harmonic, with frequency of $\Omega$ and range $\xi$; 
\begin{equation}
V_{\rm pin}=\frac{1}{2}\Mw\Omega^2(X^2-\xi^2)\theta(\xi-|X|),
\end{equation}
where $\Mw\equiv \frac{2\hbar^2 A}{\Kp \lambda a^3}
=\frac{\hbar^2 N}{\Kp\lambda^2}$ 
is wall mass, $N\equiv 2\lambda A/a^3$ is the number of spins in the wall.
In most cases, pinning range $\xi$ is of order of width, $\xi\sim \lambda$.
Pinning term in Eq. (\ref{DWeq}), $f_{\rm pin}$, is defined as 
$f_{\rm pin}\equiv \frac{\lambda}{\hbar NS} F_{\rm pin}$ with
$F_{\rm pin}\equiv -\frac{\partial V_{\rm pin}}{\partial X}$, 
 and given as
\begin{equation}
f_{\rm pin}=-\frac{\Omega^2}{2\vc}X\theta(\xi-|X|)
 = -\frac{2\Vz}{\hbar S}\frac{\lambda}{\xi^2} X\theta(\xi-|X|)
.
\end{equation}

The wall velocity in the limit of large current is obtained from Eq. (\ref{DWeq})
as
\begin{eqnarray}
\frac{\dot{X}}{\lambda} 
 &\rightarrow& 
\frac{1}{1+\alpha^2} \frac{a^3}{e\lambda}
 \left(\frac{P}{2S}+\alpha \betap \right) j \nonumber\\
 \dot{\phiz}
 &\rightarrow& 
\frac{1}{1+\alpha^2} \frac{a^3}{e\lambda}
 \left( \betap -\alpha\frac{P}{2S} \right) j.
\end{eqnarray}
Thus the spin torque efficiency, defined as 
$\eta \equiv \dot{X}\frac{2eS}{a^3 Pj}$, can be larger than unity when 
$\betap$ contribution is taken account.

\section{Numerical results}
\begin{figure}[tbp]
\begin{center}
\includegraphics[scale=1]{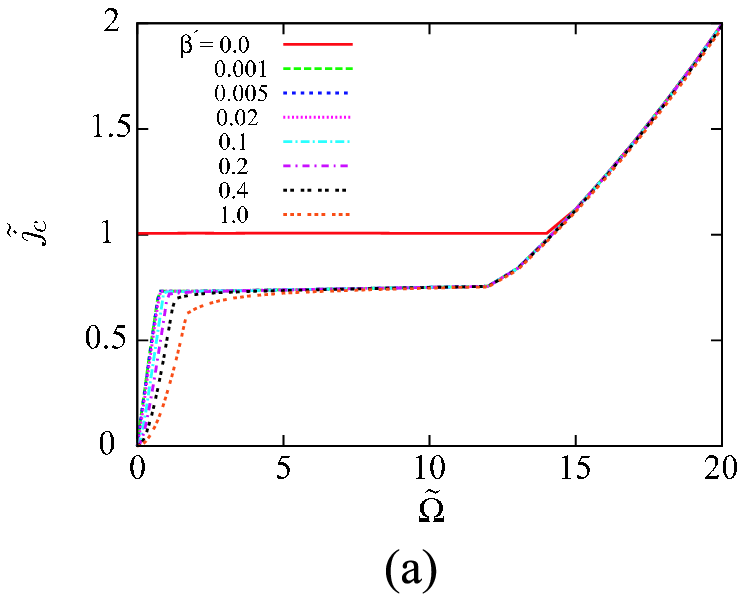}
\includegraphics[scale=1]{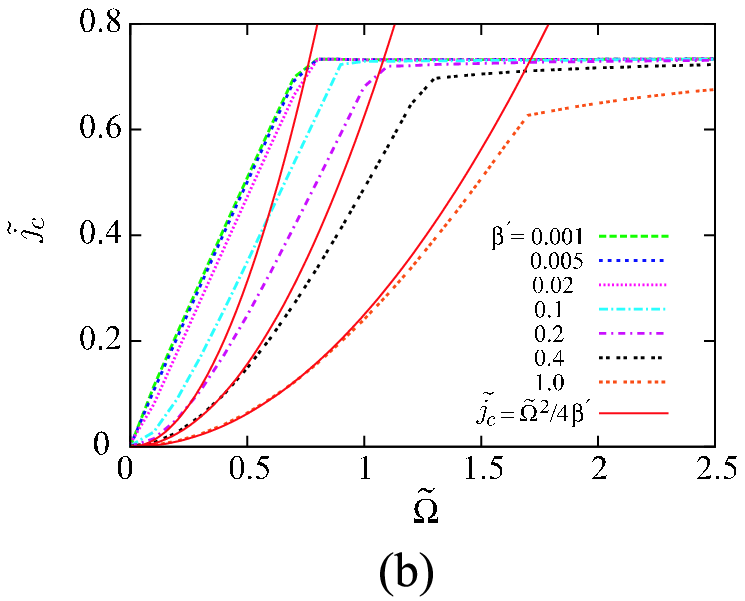}
\caption{(Color online) (a): Threshold current $\jctil$ plotted as function of pinning frequency $\Omegatil\equiv\sqrt{2\Vztil}$ for several values of $\betap$ with $\alpha=0.01$ and $\Ptil=1$. 
(b): Threshold current in weak pinning regime. 
Fitted curves for $\betap\gtrsim 0.1$ are 
$\jctil\propto\Omegatil^2$.
For small $\betap(\lesssim0.02)$, $\jctil$ is linear in $\Omegatil$.
\label{FIGjc} }
\end{center}
\end{figure}

\begin{figure}[tbp]
\begin{center}
\includegraphics[scale=1]{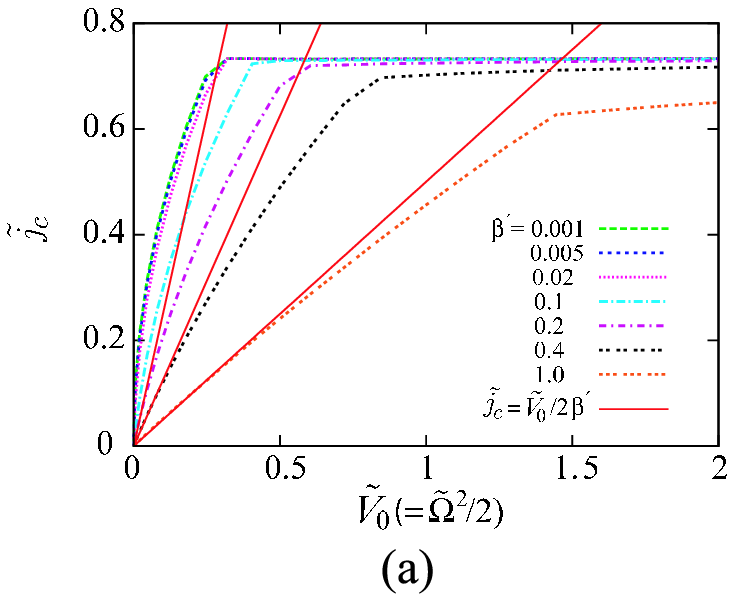}
\includegraphics[scale=1]{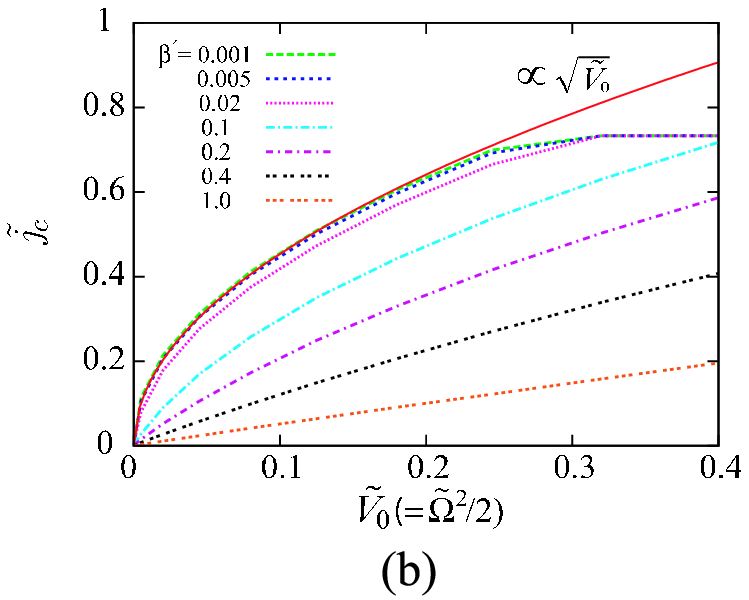}
\caption{(Color online) (a): Threshold current $\jctil$ in weak pinning regime plotted as function of pinning strength $\Vztil$ for several values of $\betap$ with $\alpha=0.01$ and $\Ptil=1$. 
For $\betap\gtrsim0.1$, $\jctil\simeq \frac{\Vztil}{2\betap}$ is linear in $\Vztil$.
(b): Weaker pinning regime. 
For small $\betap(\lesssim0.02)$, $\jctil\propto\sqrt{\Vztil}$.
\label{FIGjcV} }
\end{center}
\end{figure}

In this section, results of numerical calculation are presented.
The equation of motion solved is in terms of dimensionless parameters;
\begin{eqnarray}
\frac{\partial }{\partial \ttil} \left( {\Xtil} -\alpha {\phiz} \right)
 &=& \sin 2\phiz + \Ptil\jtil \nonumber\\
\frac{\partial }{\partial \ttil} \left( {\phiz} +\alpha{\Xtil} \right)  
       &=& -\frac{1}{2}\Omegatil^2 \Xtil
                      \theta(\frac{\xi}{\lambda}-|\Xtil|) +\betap \jtil 
    =  -\Vztil \Xtil\theta(\frac{\xi}{\lambda}-|\Xtil|) +\betap \jtil,
\label{DWeq3}
\end{eqnarray}
where
$\ttil\equiv t \vc/\lambda$, $\Xtil\equiv X/\lambda$, 
$\Omegatil\equiv \Omega \lambda/\vc 
 = \frac{2\sqrt{2}}{S}\frac{\lambda}{\xi}\sqrt{\frac{\Vz}{\Kp}}$, 
$\Ptil\equiv \frac{P}{2S}$, 
$\jtil\equiv \frac{a^3}{e\vc}j$ and 
$\Vztil\equiv \frac{1}{2}\Omegatil^2
 = \Vz \frac{2}{\hbar S} \frac{\lambda^3}{\vc \xi^2}
 = \frac{4}{S^2} \frac{\Vz}{\Kp}\left(\frac{\lambda}{\xi}\right)^2$,
 with
$\Vz\equiv \frac{\Mw}{2N}\Omega^2\xi^2
 = 
\frac{\hbar^2}{2}\frac{\Omega^2}{\Kp}\left(\frac{\xi}{\lambda}\right)^2
=\frac{\hbar S}{4}\frac{\Omega^2}{\vc}\frac{\xi^2}{\lambda}
$ 
being a pinning potential strength per spin.
Calculation is done with $\alpha=0.01$, $S=0.5$\cite{Yamaguchi05} and $P=1$. 
Threshold current is plotted as function of  $\Omegatil$ and $\Vztil$
in Figs. \ref{FIGjc} and \ref{FIGjcV} for several values of $\betap$.
It is clearly seen in Fig. \ref{FIGjc}(a) that behaviors for $\betap=0$ and 
$\betap\neq 0$ are quite different except for extremely strong pinning regime ($\Vztil \gtrsim 100\simeq 1/\alpha$).
We will come back to this difference later.

Let us focus on the case $\betap\neq 0$ first.
Fig. \ref{FIGjc} indicates that there are three regimes;
\begin{itemize}
\item[I)]
Weak pinning regime ($\Omegatil \lesssim O(1)$), $\jctil$ grows with $\Vztil$
\item[II)]
Intermediate regime ($ O(1) \lesssim \Omegatil \lesssim O({\alpha}^{-1/2}$), $\jctil\simeq $constant $\simeq 0.7-0.8$
\item[III)]
Strong pinning regime  ($\Omegatil \gtrsim O({\alpha}^{-1/2})$), $\jctil\propto\Vztil$
\end{itemize}
It should be noted here in figures that $\betap$ affects the threshold only for weak pinning, $\Vztil \lesssim 1$ 
(i.e., $\Vz\lesssim \Kp$). 
Looking closer, we see that the 
weak pinning regime actually consists of two regimes;
\begin{itemize}
\item[I-a)]
Small $\betap/\Omegatil$: $\jctil\propto \sqrt{\Vztil}$
\item[I-b)]
Large $\betap/\Omegatil$: $\jctil\propto {\Vztil}/\betap$
\end{itemize}

Physical picture of these regimes is as follows.
In regime I, current is low, so $\phiz$ does not grow, and 
dynamics is described by $X$. 
In regime I-a), $\betap$ is negligible and depinning is due to spin-transfer.
Namely, spin-transfer term (proportional to $\Ptil$) gives a finite velocity to domain wall and the wall escapes from the pinning potential 
by use of initial kinetic energy supplied from the spin-transfer torque.
In other words depinning is due to a kinetic energy supplied by spin transfer.
In regime I-b), depinning is governed by a force. It occurs when the force due to current exceeds pinning force, as is seen from the expression $\jctil\propto  {\Vztil}/\betap$.
Dynamics in regime II) and III) is described by $\phiz$. 
Regime II) is a regime where depinning is due to spin torque, but terminal velocity is determined by $\betap$. 
Depinning mechanism here is the same as the intrinsic pinning pointed out in ref. \cite{TK04}, but the lowering of $\jc$ by a factor of $0.7\sim0.8$ by $\betap$-term is one of new results of the present paper. 
This is the reason why $\jc$ looks different for $\betap=0$ and $\betap\neq 0$ in Fig. \ref{FIGjc} (a).
Regime III) is a strong pinning regime argued in ref. \cite{TK04}. Depinning here is due to spin-transfer torque.
In both regimes II) and III), $\betap$ does not affect the threshold current.

Let us look into each regime closely.

\section{Analytical results}

\subsection{Weak pinning regime : I)}

Under weak current, $\jtil\lesssim 1$, $\phiz$ remains small and wall dynamics is well described by $X$ only. 
This is the first regime I).
Linearizing sine-term in Eq. (\ref{DWeq}) as $\sin 2\phi \simeq 2\phi$, $\phi$ can be eliminated to obtain a simple equation for $X$ as\cite{SMYT04,TSIK05}
\begin{equation}
(1+\alpha^2)\partial_{\ttil}^2 {\Xtil}
 +\frac{1}{\tautil}\partial_{\ttil}{\Xtil}
+\Omegatil^2 \Xtil = \tilde{F}_{\rm j},
\label{DWeq2}
\end{equation}
where 
$1/\tautil=2\alpha\left(1+\frac{1}{2}\Vztil\right)$
and 
\begin{equation}
\tilde{F}_{\rm j}\equiv 2\betap \jtil,
\end{equation}
is a dimensionless force due to current.
We consider a case of steady current and weak damping; $2\Omegatil\tautil > 1$
A general solution to the equation (\ref{DWeq2})  
(inside the pinning potential) is given as
\begin{equation}
\Xtil(t)=  \frac{\betap \jtil}{\Omegatil^2}
+e^{-\frac{\ttil}{2\tautil}}
 (A\cos\Omegaptil \ttil 
 + B\sin \Omegaptil \ttil )  ,
\end{equation}
where 
$\Omegaptil\equiv \sqrt{\Omegatil^2-\frac{1}{4\tautil^2}}$
and $A$, $B$ are constants.
Initial condition required is $\Xtil(0)=0$
and $\partial_{\ttil}\Xtil(0)=\Ptil \jtil$.
The second condition on the wall speed comes from the first equation in Eq. (\ref{DWeq3}) (with $\phiz(0)=0$), and is the most important
consequence of spin-transfer torque; namely, spin-transfer torque gives initial speed to the wall. 
With these initial conditions, we obtain
\begin{equation}
\Xtil(t)=  \frac{2\betap \jtil}{\Omegatil^2}
\left(1-e^{-\frac{\ttil}{2\tautil}}
 \left(\cos\Omegaptil \ttil 
 +\frac{1}{2\Omegaptil \tautil} \sin \Omegaptil \ttil \right) \right) 
+\frac{\Ptil \jtil}{\Omegaptil} 
e^{-\frac{\ttil}{2\tautil}} \sin \Omegaptil \ttil.
\end{equation}
The first part is governed by a force from $\betap$ and the second is driven by a spin-transfer torque term.
In the case of small $\betap$, spin torque contribution leads to 
a maxmum displacement 
\begin{equation}
|{\Xtil}_{\rm max}|\simeq \frac{\Ptil}{\Omegaptil}\jtil,
\end{equation}
while
\begin{equation}
|{\Xtil}_{\rm max}|\simeq \frac{4\betap}{\Omegatil^2}\jtil,
\end{equation}
if $\betap$ is large. 
(We assumed here that damping is weak ($\Omegap\tau\gg1$).) 
The first regime corresponds to regime I-a) and the second to I-b).
Threshold current in each case is given as
\begin{equation}
{\jctil}^{\rm Ia)} \sim \frac{\Omegatil}{\Ptil} \frac{\xi}{\lambda},
\end{equation}
and
\begin{equation}
{\jctil}^{\rm Ib)} \sim \frac{\Omegatil^2}{4\betap} \frac{\xi}{\lambda}.
\end{equation}
The crossover occurs at
\begin{equation}
\betap_{\rm c}\simeq \frac{\Ptil}{2}\Omegatil.
\end{equation}
In terms of dimensionful quantities, 
\begin{equation}
{\jc}^{\rm Ia)} \simeq  \frac{2\sqrt{2}S}{P} \frac{e}{a^3}\frac{\sqrt{\Kp\Vz}}{\hbar} \lambda
=\frac{2\sqrt{2}}{S}\sqrt{\frac{\Vz}{\Kp}} \jci,\label{jcIa}
\end{equation}
and
\begin{equation}
{\jc}^{\rm Ib)}= 
\frac{e\lambda}{4 \vc a^3}
\frac{\Omega^2}{|\betap|}\xi 
 = \frac{Se\Vz}{\hbar a^3} 
\frac{1}{|\betap|} \frac{\lambda^2}{\xi}
 =\frac{P}{S} \frac{1}{|\betap|}\frac{\Vz}{\Kp} \frac{\lambda}{\xi} \jci.
 \label{jcIb}
\end{equation}
Note that simple comparison of pinning force and $F_{\rm j}$ in Eq. (\ref{DWeq2}) gives a result correct up to a numerical factor, 
${\jc}^{\rm Ib)}\sim \frac{1}{2}\frac{Se\Vz}{\hbar a^3} 
\frac{1}{|\betap|} \frac{\lambda^2}{\xi}$.

The pinning strength $\Vz$ is experimentally accessible by driving the wall by magnetic field.
Magnetic field $B$ along easy axis add a term in Eq. (\ref{DWeq})
\begin{equation}
f_{B} = \frac{g \muB}{\hbar}B, \label{fb}
\end{equation}
where $g=2$.
By a simple comparison of pinning force and magnetic field, $\Vz$ is written in terms of the  depinning magnetic field $\Bc$ as
\begin{equation}
\Vz= \frac{S}{2}g\muB \Bc \frac{\xi}{\lambda}, \label{VzB}
\end{equation}
 and so ${\jc}^{\rm Ib)}$ is simplified to be
\begin{equation}
{\jc}^{\rm Ib)} = \frac{e }{\hbar a^3}\frac{S^2}{2}g \muB \Bc \lambda
   \frac{1}{|\betap|}  .
 \label{JcIb}
\end{equation}

\subsection{Intermediate regime : II (Intrinsic pinning)}
This regime could be important for application since the threshold is not sensitive to the sample irregularities.
Depinning in this regime  
$\jtil\gtrsim O(1)$ is described by $\phiz$ as done in ref. \cite{TK04}.
The reason is that the effective mass of $\phiz$-"particle", given by $1/\Vz$\cite{TT96}(see Eq. (\ref{phieq})), becomes lighter than the corresponding mass of $X$-"particle" given by $1/\Kp$, and so $\phiz$-"particle" is a better variable to describe dynamics for strong pinning. 
By eliminating $X$ from Eqs. (\ref{DWeq3}), we obtain
\begin{equation}
(1+\alpha^2) \partial_{\ttil}^2 \phiz
+ \alpha\partial_{\ttil}{\phiz}
\left(2\cos2\phiz +\Vztil\right)
+\Vztil\sin2\phiz + \jtil\Vztil\Ptil=0.
\label{phieq}
\end{equation}
Thus $\betap$ does not affect the dynamics of $\phiz$.
(Correctly, this feature is specific to a harmonic pinning potential, and anharmonicity results in appearance of $\betap$.
In fact, $\betap$-term is eliminated from the equation of motion if one replaces $X$ in Eq. (\ref{DWeq3}) by 
$X'\equiv X-\frac{2\betap}{\Omegatil^2}\jtil$ (i.e., shift of stable point of $X$).
Even in anharmonic case, nevertheless, we have numerically checked that the $\betap$ does not lead to important contribution in this regime.)

From Eq. (\ref{phieq}), we see that the energy barrier for $\phiz$ vanishes when
$
\jctil\sim \Ptil^{-1}, 
$ 
irrespective of pinning strength.
Once $\phiz$ escapes from local minimum, its velocity is given by Eq. (\ref{phieq}) as
\begin{equation}
\partial_{\ttil}\phiz \simeq \frac{\jtil\Ptil}{\alpha}.
\end{equation} 
This corresponds by use of Eq. (\ref{DWeq3}) 
to a maxmum displacement of the wall of 
\begin{equation}
\Xtil_{\rm max} \simeq - \frac{1}{\Vztil} \partial_{\ttil}\phiz
 \simeq \frac{\jtil\Ptil}{\alpha\Vztil}.
\label{XII}
\end{equation}
For intermediate pinning strength, $\alpha\Vztil \lesssim 1$, 
$|\Xtil_{\rm max}|$ exceeds $\xi/\lambda$ (if $\xi/\lambda\sim1$), i.e., depinning of $X$ 
occurs as soon as $\phiz$ is depinned.
Thus the threshold is roughly given by $\jctil\sim \Ptil^{-1}$, but this estimate turns out to be too rough to determine numerical factor correctly.
In fact, numerical result in Fig. \ref{FIGjc} indicates that $\jctil$ is actually given by
\begin{equation}
\jctil\sim 0.7\times \Ptil^{-1} \;\;\; (\betap\neq0),\label{jcII}
\end{equation}
while $\jctil\sim \Ptil^{-1}$ when $\betap=0$.
Result of $\betap=0$ case is in agreement with ref. \cite{TK04}.
The reason for the difference between $\betap=0$ and $\betap\neq0$ 
is understood as follows. 
In the case of $\betap=0$, 
even if $X$ escapes from the pinning center at current 
$\jtil > 0.7\Ptil^{-1}$,
terminal velocity vanishes if $\jtil <\Ptil^{-1}$, since the motion stops
due to the intrinsic pinning effect (i.e., $\phiz$ reaches a steady value and $\dot{X}$ becomes zero).
On the other hand, if $\betap\neq 0$, steady motion of $X$ is possible as soon as $X$ escapes from the pinning ($\jtil > 0.7\Ptil^{-1}$).
This is the reason the threshold value in intermediate regime is different for $\betap=0$ and $\betap\neq0$
(Fig. \ref{FIGjc}).

If $\beta' < 0$, or more precisely, if the relative sign between 
$\beta'$ and $\tilde P$ in Eq.(\ref{DWeq3}) is negative, the $\beta'$-term 
will drive the depinned wall back to the pinning center, and the 
threshold in this regime is given by the intreinsic value $\jctil=1$. 

\subsection{Strong pinning regime : III}
Eq. (\ref{XII}) indicates that for extremely strong pinning, 
$\Vztil \gtrsim \alpha^{-1}$, the wall is not always depinned even after $\phiz$ escapes from the potential minimum. 
Depinning occurs at
\begin{equation}
\jctil\sim \frac{\alpha\Vztil}{\Ptil}\frac{\xi}{\lambda},
\label{jcIII}
\end{equation}
as has been pointed out in ref. \cite{TK04}.

\section{Wall speed}
\subsection{Close to threshold}
Another important quantity to understand mechanism of wall motion is the wall velocity at $\jc$.
We consider a steady current or a pulse with duration long enough 
(longer than time scale of $\tau$), and the weak pinning case, $\jc \ll \jci$ (i.e., regime I).
The terminal wall velocity after depinning ($j>\jc$)  is obtained from eq. (\ref{DWeq2}) with $\Omega=0$ as 
\begin{equation}
\tilde{v}\equiv \partial_{\ttil} {\Xtil}= \frac{\betap}{\alpha}\jtil.
\end{equation}
At $\jc$, the wall velocity suddenly jumps from zero to $\frac{\betap}{\alpha}\jc$, 
and so the jump in dimensionless unit is given by
$\Delta \tilde{v}^{\rm Ia)} 
=\frac{\betap}{\alpha}\frac{\Omegatil}{\Ptil}\frac{\xi}{\lambda}$ in regime I-a) and
$\Delta \tilde{v}^{\rm Ib)} 
=\frac{1}{\alpha}\frac{\Omegatil^2}{4}\frac{\xi}{\lambda}$ in regime I-b).
In terms of dimensionful quantity, 
$\Delta {v}/\jc=\frac{\betap}{\alpha}\frac{a^3}{e}$, we have 
\begin{eqnarray}
\Delta {v}^{\rm Ia)} &=& 
\frac{\betap}{\alpha}\frac{2\sqrt{2}S}{P}\frac{\sqrt{\Kp\Vz}}{\hbar}
\lambda \nonumber\\
\Delta {v}^{\rm Ib)} &=& 
\frac{1}{\alpha}\frac{S\Vz}{\hbar}\frac{\lambda^2}{\xi}.
\end{eqnarray}
These behaviors are seen in Fig. \ref{FIGvj}.
Comparing with result of ref. \cite{Thiaville05}, behavior appears qualitatively similar in the large $\betap$ regime.
Detailed comparison is, however, not possible since value of $\Kp$ and $\Vz$ in ref. \cite{Thiaville05} are not known.
It should be noted that the wall velocity near extrinsic threshold is discontinuous at zero temperature. 
This is because the wall as soon as depinned feels a tilted potential (due to $\betap$) and has finite velocity.
This might explain rather scattered experimental data for velocities in metals\cite{Klaeui03,Yamaguchi04}.
In contrast, results on semiconductor indicates quite smooth temperature dependence\cite{Yamanouchi06}, which might suggest different origins of threshold from those in metals.

\subsection{After depinning (or no pinning)}
Let us see how wall velocity behaves after the wall is depinned.
We solve here Eq. (\ref{DWeq3}) with $\Vztil=0$, which is a simple equation of $\phi$ given as
\begin{equation}
{\partial_\ttil} {\phiz} = \frac{1}{1+\alpha^2}  
 \left( (\betap-\Ptil\alpha)\jtil-\alpha \sin 2\phiz \right) .
\end{equation}
For a steady current, this equation has a solution of
\begin{equation}
\sin 2\phiz = 
\left( \frac{\betap}{\alpha}-\Ptil \right) \jtil
-\sgn(\Ptil-\frac{\betap}{\alpha}\jtil)
\frac{\sqrt{\left[\left(\Ptil-\frac{\betap}{\alpha}\right)\jtil\right]^2-1}}
{|(\betap-\alpha\Ptil)\jtil| +\alpha \sin(2\om t-\vartheta)}
,
\label{sin2phi}
\end{equation}
where 
$\om \equiv \frac{\alpha}{1+\alpha^2}
\sqrt{\left[\left(\Ptil-\frac{\betap}{\alpha}\right)\jtil\right]^2-1}$, 
$\sin \vartheta\equiv\frac{1}{|\Ptil-\frac{\betap}{\alpha}\jtil|}$, 
and $\sgn(B)$ denotes sign of $B$.
The wall velocity is given from Eq. (\ref{DWeq3}) as
\begin{equation}
{\partial_\ttil} {\Xtil} = \frac{1}{1+\alpha^2}  
 \left( \sin 2\phiz + (\Ptil+\alpha\betap)\jtil \right). 
 \label{DWvelocity}
\end{equation}
We see that anomaly in the velocity appears when $\om$ switches from real to imaginary, i.e., at $\jtil = \jatil$, where 
\begin{equation}
\jatil \equiv \frac{1}{\left|\Ptil-\left(\frac{\betap}{\alpha}\right)\right|}.
\end{equation}
Above and below $\jatil$, wall dynamics is quite different. 
Above $\jatil$, wall velocity (\ref{DWvelocity}) has an oscillating component, while wall reaches a steady motion if below $\jatil$.
The average velocity for $\jtil \geq \jatil$ is obtained by use of 
Eqs. (\ref{sin2phi}) (\ref{DWvelocity}) and 
\begin{equation}
\frac{1}{T} \int_0^T dt \frac{1}{A\sin \left(2\pi \frac{t}{T}\right) +B}
 =\frac{1}{\sqrt{B^2-A^2}},
\end{equation}
(for $|B|>A$ and $T=2\pi/\om$) to be
\begin{equation}
\average{\dot{X}} = \frac{\betap}{\alpha} \jtil
 +\frac{\sgn[(\Ptil-\frac{\betap}{\alpha})\jtil]}{1+\alpha^2}
\sqrt{\left[\left(\Ptil-\frac{\betap}{\alpha}\right)\jtil\right]^2-1}.
\end{equation}

\begin{figure}[tbp]
\begin{center}
\includegraphics[scale=1]{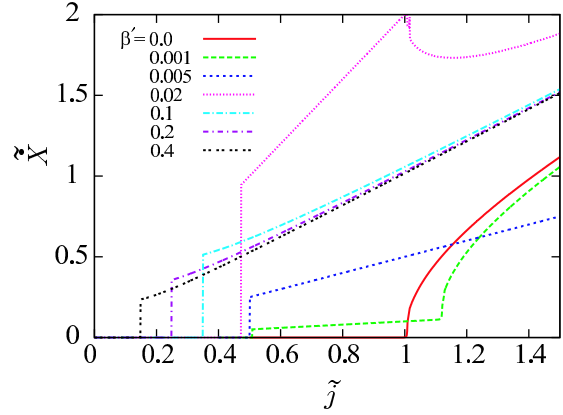}
\caption{(Color online) Wall velocity as function of current for $\Omegatil=0.5$ and $\alpha=0.01$.
A jump in wall velocity is seen at $\jtil=\jctil$.
Crossover from I-a)($\jctil\sim \Omegatil/\betap$) to 
I-b)($\jctil \sim \Omegatil$) regime is seen at 
$\betap\simeq \frac{\Omegatil}{4}\sim 0.1$. 
\label{FIGvj} }
\end{center}
\end{figure}

Below $\jatil$, $\sin2\phiz$ approaches at $t\rightarrow\infty$ a steady value of 
\begin{equation}
\sin 2\phiz \rightarrow \left(\frac{\betap}{\alpha}-\Ptil\right) \jtil,
\end{equation}
and so terminal velocity is obtained as
\begin{equation}
\average{\dot{X}} = \frac{\betap}{\alpha} \jtil.
\end{equation}
Note that the analysis here is the case of $\Vztil=0$. 
In the presence of extrinsic pinning, velocity vanishes below $\jctil$ 
as seen in Fig. \ref{FIGvj}.

The anomaly at $\jatil$ are seen in Fig. \ref{FIGvj} in the case of 
$\betap=0.001$ (at $\jtil=\frac{1}{0.9}\sim 1.1$)
and 
$\betap=0.02$ (at $\jtil=1$), and the behavior is in agreement with 
numerical result of ref. \cite{Thiaville05}, where 
the anomaly of velocity under current was first reported.
This anomaly is essentially the same as the anomaly under magnetic field, known as Walker breakdown\cite{Schryer74,Hubert98}.

Let us here briefly consider the effect of external magnetic field along easy axis.
As seen from in Eq. (\ref{fb}), the field replaces
$\betap \jtil$ by $\betap \jtil +b$, where
$b\equiv -\frac{g\muB B_z\lambda }{\hbar \vc} = 
-\frac{2g \muB B_z}{S\Kp}$ ($g=2$).
The average wall velocity with $\Vztil=0$ is then given 
if $\jtil \geq \jatil$ by
\begin{equation}
\average{\dot{X}} \rightarrow \frac{\betap\jtil+b}{\alpha} 
+\frac{\sgn[[(\Ptil-\frac{\betap}{\alpha})\jtil]- \frac{b}{\alpha}]}
{1+\alpha^2}
\sqrt{\left[\left(\Ptil-\frac{\betap}{\alpha}\right)\jtil- \frac{b}{\alpha}\right]^2-1},
\end{equation}
 and by
\begin{equation}
\average{\dot{X}} \rightarrow \frac{\betap \jtil+b}{\alpha} ,
\end{equation}
if $\jtil < \jatil$,
where the anomaly now occurs at
\begin{equation}
\jatil \rightarrow \frac{b\pm \alpha}
{\Ptil \alpha-{\betap} },
\end{equation}
the $\pm$ denotes $\sgn[(\Ptil \alpha-{\betap})-b/\jatil]$.

\section{Discussion}
\begin{table}[tb]
\caption{Summary of threshold current. The second to the last column indicates the mechanism of depinning, either spin transfer or force ($\betap$-term).
The last column is a "good" variable to describe depinning.
We see that purely intrinsic pinning occurs in regime II, while 
other regimes are affected by extrinsic pinning.
\label{table}}
\begin{center}
\begin{tabular}{|c|c|c|c|c|} \hline
  regime   &     & threshold & depinning mechanism &  \\ \hline
I-a: Weak pinning & $\Omegatil \lesssim O(1)$, $\betap\lesssim O(\Omegatil)$ & $\jc\propto \sqrt{\Kp\Vz}$ & spin transfer & $X$ \\ \hline
I-b: Weak pinning & $\Omegatil \lesssim O(1)$, $\betap\gtrsim O(\Omegatil)$ & $\jc\propto {\Vz}/\betap$ & $\betap$ & $X$ \\ \hline
II: Intermediate pinning & $O(1) \lesssim \Omegatil \lesssim O(\alpha^{-1})$ & $\jc\propto \Kp$ & spin transfer & $\phiz$ \\ \hline
III: Strong pinning &  $O(\alpha^{-1}) \lesssim \Omegatil$  & $\jc\propto {\Vz}/\alpha$ & spin transfer & $\phiz$ \\ \hline
\end{tabular}
\end{center}
\end{table}
Let us summarize the results in table \ref{table}.
It is interesting that such a simple set of equation of motion results in so rich behaviors.
This fact is very important for device application, since
reduction of threshold current, which is a must to develop MRAM based on current-driven domain wall motion, can be realized in different ways depending on system. 
For instance, in regime II), threshold is governed by the average sample shape, and not by the sample quality (roughness) as pointed out in ref. \cite{TK04}. 
In contrast, region I-b) is most strongly affected since magnetic imperfection results in $\Vz$, and magnetic defects modifies $\betap$ strongly by affecting the electron transport.
Regime I-a) is moderately affected by both sample shape ($\Kp$) and  
 defects ($\Vz$).
Existence of these different threshold may explain the controversy among experimental results. 
In fact, some experiments indicate threshold is very insensitive to artificially introduced pinning centers\cite{Parkin04}, while
some experiments indicates pinning is essential\cite{Himeno05}.

So far experimental results on metallic samples all showed that threshold current is of order of $10^{12}$[A/m$^2$].
If we use experimentally estimated value of $\Kp\sim O(1)$[K]\cite{Yamaguchi05}, 
the observed threshold is orders of magnitude ($10^{-2}-10^{-1}$ times) smaller than the intrinsic threshold, $\jci$.
For instance, a sample of Yamaguchi\cite{Yamaguchi04,Yamaguchi05} showed 
$\jc=1\times 10^{12}$[A/m$^2$].
The anisotropy energy is estimated to be $\Kp=2.4$[K], and using
$S\sim \frac{1}{2}$, $a\simeq 2.2$\AA\ and $P\sim O(1)$, we obtain 
$\jci=5.8\times 10^{13}$[A/m$^2$], i.e., 
$\jc/\jci\sim 0.02$.
The observed low threshold in metals thus should be regarded as due to an extrinsic pinning in regime I-a) or I-b).
Let us first try to explain experimental result\cite{Yamaguchi05} assuming regime I-a).
We assume $\xi\sim \lambda$.
Pinning potential is estimated from the measured depinning field of
$\Bc=0.01-0.1$[T].
By use of Eq. (\ref{VzB}), 
$\Vz= 0.34\times (10^{-2} \sim 10^{-1})$[K]
$=4.7\times (10^{-26} \sim 10^{-25})$[J], i.e.,
$\frac{\Vz}{\Kp} = 1.4\times (10^{-3} \sim 10^{-2})$.
Thus
${\jc}^{\rm Ia)} = (0.21\sim0.67)\times \jci$.
This value is still too big to explain the experimental value.
Velocity jump becomes 
$\Delta v^{\rm Ia)} =\frac{\betap}{\alpha}\times 839$[m/s], so extremely small $\betap$ ($\frac{\betap}{\alpha}\sim 4\times 10^{-3}$) is required to explain experimental value of $\Delta v\sim 3$[m/s]\cite{Yamaguchi04}.
If we assume regime I-b), the threshold is
${\jc}^{\rm Ib)} = \frac{1}{|\betap|} \times 
 2.8 \times (10^{-3} \sim 10^{-2}) \times \jci$.
Experimental value could be reproduced if $\betap= 0.1 \sim 1$.
But such large value of $\betap$ cannot be explained within the current understanding that $\betap$ arises from either non-adiabaticity\cite{TK04,Thiaville05} or spin relaxation\cite{Zhang04}.
Instead, $\Delta v$ cannot be explained by use of the above $\Vz$ assuming I-b), as it predicts too large value of 
$\Delta v^{\rm Ib)}=10^3$[m/s].
Thus, honestly, none of the above predictions are successful in explaining experimental result of metals quantitatively.

There remain some possibilities to resolve this disagreement. For instance, estimate of $\Vz$ by use of experimental $B_c$ could be an over-estimation
if effective barrier height $\Vz$ is greatly reduced by heating under current, while such heating does not occur under static magnetic field.
Let us estimate the pinning potential which gives the experimental 
value of $\jc$.
Assuming regime I-a), experimental value of $\jc/\jci=0.02$ is reproduced if
$\mu\equiv \frac{\Vz}{\Kp}=1.3\times 10^{-5}$, which corresponds to 
$\Vz=3\times 10^{-5}$[K]$=4.5\times 10^{-5}$[T].
This is two orders of magnitude smaller than the value extracted from $B_{\rm c}$.
For I-b), we have $\mu=\betap\times 10^{-2}$.
From the experiment, 
$\Delta v /(\frac{a^3}{e}\jc) = 3$[m/s]$/67$[m/s]$=0.05$. 
This value is equal, for regime I, to $\frac{\betap}{\alpha}$, so
$\betap=5\times 10^{-4}$ if $\alpha=0.01$.
So in case I-b), $\mu=5\times 10^{-6}$.
Thus, assuming either regime I-a) or I-b), the experimental results could be explained by an extremely weak pinning potential,
$\frac{\Vz}{\Kp}=10^{-6}\sim 10^{-5}$.

The situation is very different for magnetic semiconductor\cite{Yamanouchi06}.
The observed $\jc$ is in quantitative agreement with intrinsic threshold, $\jci$, and the velocity above threshold is also in agreement with spin-transfer mechanism. 
Thus extrinsic pinning and $\betap$ does not seem to play roles. 

Identification of the origin of threshold is thus necessary to lower the threshold current in experiments.
It would be very important first to see  if thresholds observed in metallic systems are more or less governed by $\Vz$ and $\betap$, and second to try to lower threshold by controlling sample quality ($\Vz$) and $\betap$ by 
introducing heavy-atom impurities.

The authors are grateful to Y. Yamaguchi, T. Ono, M. Yamanouchi, H. Ohno, M. Kl\"aui, A. Thiaville, J. Ieda, N. Nagaosa, J. Inoue, S. Maekawa and S. Parkin for valuable discussion.


\end{document}